\documentclass[conference]{IEEEtran}
\IEEEoverridecommandlockouts
\pdfoutput=1 
\usepackage{cite}
\usepackage{amsmath,amssymb,amsfonts}
\usepackage{algorithmic}
\usepackage{graphicx}
\usepackage{textcomp}
\usepackage{xcolor}
\usepackage{float}
\usepackage{hyperref}
\usepackage{soul}

\def\BibTeX{{\rm B\kern-.05em{\sc i\kern-.025em b}\kern-.08em
    T\kern-.1667em\lower.7ex\hbox{E}\kern-.125emX}}

\makeatletter
\newcommand{\linebreakand}{%
  \end{@IEEEauthorhalign}
  \hfill\mbox{}\par
  \mbox{}\hfill\begin{@IEEEauthorhalign}
}
\makeatother

\begin{document}

%Introduction - cs education, engagement, motivation. Tell what this is and why it's important.
%Background - will need background on service learning research

\title{Iterative Service-Learning: A Computing-Based Case-study Applied to Small Rural Organizations}
%Iterative Service Learning As a Means to Improve Social Media Marketing for Small Rural Organizations

\author
 {\IEEEauthorblockN{Sherri Weitl-Harms}
 \IEEEauthorblockA{\textit{Computer Science, Design, and Journalism} \\
 \textit{Creigton University}\\
 Omaha, NE USA \\
 sherriweitlharms@creighton.edu}
 }

\maketitle

%up to 8 pages plus 1 page for references

\begin{abstract}
This innovative practice full paper describes the iterative use of service learning to develop, review, and 
improve computing-based artifacts for small rural organizations, over an extended period. It is well-
known that computing students benefit from service-learning experiences as do the community 
partners. It is also well-known that computing artifacts rarely function well long-term without versioning
and updates. Service-learning projects are often one-time engagements, completed by single teams of
students over the course of a semester or year long course. This limits the benefit for the community
partners, such as small rural organizations, that do not have the expertise or resources to review and
update a project on their own.

Over the course of several years, teams of undergraduate students in a computing capstone social media development course created tailored social media plans for numerous small rural organizations. The projects were required to meet the client's specific needs, with identified audiences, measurable goals, and a minimum of three recommended social media strategies and tactics to reach the identified goals. This paper builds on previously reported initial results for 60 projects conducted over several years.  Nine clients were selected to participate in the iterative follow-up process, where new student teams conducted client interviews, reviewed the initial plans, and analyzed metrics from the social media strategies and tactics already in place to provide updated, improved artifacts. Using ABET computing learning objectives as a basis, clients reviewed the student teams and the artifacts created. Students also reflected on their experiences. %Overwhelmingly, both the client and student feedback from this approach were positive. 
This research provides a longitudinal study of the impact of the interventions in increasing implementation and sustained use rates of computing artifacts developed through service learning, along with lessons learned. Both students and clients reported high satisfaction levels, and clients were particularly satisfied with the iterative improvement process. This research demonstrates an innovative practice for creating and maintaining computing artifacts through iterative service learning, while addressing the resource constraints of small rural organizations.

\end{abstract}


\section{Introduction and Background}
\subsection{Service Learning in Computing and Engineering}

Service learning is \emph{an approach to teaching and
learning in which students use academic and civic knowledge
and skills to address genuine community needs} \cite{NSLC}. Through service learning in computing and engineering courses, students collaborate with community partners, and apply their classroom knowledge to real-world situations \cite{Homkes,WEITLHARMS2024USI}. Students develop skills, while contributing to participatory community development, and receive education in values \cite{Sandoval}.  %Students in service learning courses actively co-construct their learning experiences in authentic situations, which is considered a “high impact practice” for retaining students [31]. Rather than passive consumers of instruction, students spend substantial time on learning activities and gain confidence in integrating their knowledge and skills into real-world situations [18]. Deep interaction and involvement in social and academic processes and practices on campus are shown to result in increased sense of belonging, which is predictive of undergraduate student persistence\cite{Barker}
As computer scientist Michael Goldweber, an advocate for “computing for the social good” notes\cite{Goldweber}: %argues, \emph{the modern university (and the American liberal arts college in particular) was created in theocratic societies to give students a “liberating education” that allowed them to serve a greater “common good.”} Goldweber 
 \emph{It is imperative to give computer science students as part of their professional development, the opportunity to address the following fundamental question: “Instead of using computers to make money or do basic scientific research, how can one use computing to do good in the world?”} Service learning provides students with an opportunity to recognize and respect differences and understand their responsibility in relation to others and to the common good from a technical and professional perspective\cite{Khan, Kilkenny}.

%Several students made explicit connections between the service-learning project and engineers’ responsibility to the public. Students attributed these realizations to participating in the service-learning project. They articulate in their written reflections that the project helped them realize that serving the community is an important part of the obligations of engineers. Service learning, according to the students, can heighten their altruism and desire to engage with the public, but it does not seem to be enough to move individual students beyond absences from empathy, or apathy, to the desired prosocial helping behaviours. \cite{Patterson}

Engineering and computing programs have been slow to adopt service learning as a pedagogy, but there has been a steady increase in examples of courses and programs over the last two decades, %especially in engineering 
\cite{Oakes,Riaji}. Some of the engineering programs include computing, such as the EPICS Program at Purdue University \cite{Coyle, Linos} and Butler University \cite{JohnsonL}. A systematic review of service learning experiences in computer and information science (CIS) was conducted in \cite{Barker}. They described results from 84 published works that reported service learning in higher education CIS programs.  They found 49 examples of software development service learning publications, include designing a guided tour using a mobile app, developing databases to help the community partner track scheduling or inventory information, designing websites, or developing educational games. Another systematic mapping study \cite{Baltierra} presented a review of 15 research studies that describe service learning focused computer engineering majors \cite{Baltierra}. Other computing examples include giving students an opportunity to work with a regional Native American organization to create a database solution for a maple syrup operation \cite{Khan};  having students each self-select a small database project for a real client \cite{Weitl-Harms}; and  having student teams provide social media plans for small rural organizations \cite{WEITLHARMS2024USI}.%; and a project in Malaysia for rural communities \cite{Sulaiman}.

%In computing, there are fewer examples published and little research on the impact of the pedagogy \cite{Oakes}. 
Hundreds of studies highlight the student benefits of participating in well-designed service learning courses for meeting academic learning objectives, acquiring professional skills, and improving civic engagement \cite{Barker,Riaji, Harrell}.
%Filling the gap between the industry needs and graduated students’ competencies increases employability.
Researchers have identified numerous benefits to service learning in computing programs,  including attracting and retaining students; preparing students for the workforce by developing professional skills; meeting curricular guidelines; and providing a technical service for resource-constrained organizations \cite{Robledo, Shafaat, Kilkenny}. Service learning offers a project-based experience that can show students the types of impact they can have with their major and can teach them about ethical and moral dilemmas that often arise in technical fields \cite{Robledo,Patterson}.

\subsection{Implementation of Service Learning in Computing}
Capstone courses are a common place to implement service-learning \cite{murphy,Clua}. Students are motivated when they have an opportunity to be involved in a hands-on realistic, interdisciplinary project addressing a real community need \cite{Clua}. Many of the computing service learning projects involve the creation of software or web sites, which were most commonly assessed by evaluating the tangible artifact that was produced, through presentations, reflection essays, written reports, and/or peer evaluations \cite{Barker, Oakes}. 

Incorporating service-learning projects requires extra time and organization from the instructor (e.g., soliciting non-profit organizations to serve as community partners, ensuring the partners are satisfied with their collaboration with the students), as well finding some method of providing maintenance and technical support once the course is over \cite{Chao}. 
 
Setting up successful collaborations between campuses and community partners can be hard \cite{Medero}.  One approach is to use repeat community partners, who are willing to repeatedly work with students on projects \cite{Medero}, such as through sponsoring proprietary client-oriented software projects \cite{MacKellar}.  However, these projects are often standalone, one-off projects since companies may be reluctant to have students work on their internal codebase, or to develop mission critical software; which often means that it is often difficult to get time and attention from personnel at the company during the project \cite{MacKellar}. Additionally, these types of projects often do not truly fit the definition of "service learning"  of doing "computing for social good". Another approach is to engage students via a relatively small project that fits in a one semester course, with a local nonprofit organization as the client. Local nonprofits are often happy to collaborate on these projects since they may have needs for mission-critical software systems that are not well met by the commercial software industry, yet they have limited technology budgets \cite{MacKellar}.

One well developed model of computing-based service learning that demonstrates legitimately useful software products for non-profit clients is Engineering Projects in Community Service (EPICS)\cite{JohnsonL}. The same client is often engaged with several times, but the results of each engagement are complete and not dependent on future work; while future engagements are often build on the past \cite{Mertz}. Client and student expectations are communicated early, and teams are expected to meet regularly \cite{JohnsonL}. This model works well at large universities that have resources for holistic project management and client engagement.

%When clients are onboarded as new EPICS participants, it is made clear that they are expected to commit the necessary time to mentor our students on a regular basis. To facilitate such mentorship, teams are expected to schedule regular meetings with the client both on campus and at the client site (if applicable) We recognized the program’s potential to create rewarding student experiences and strong resume building activities such as the cultivation of real-world technical expertise and the accomplishment of a goal via team driven methodologies. We showcased how these efforts have given birth to \cite{JohnsonL}

%\subsection{Computing Artifact Handoff}

Handing off the project to the community partner is a key component in service learning. Scaffolded Projects for the Social Good (SPSG) is an adaptable service-learning framework \cite{Kurkovsky2024}, in which the development phase is structured using the agile methodology. The transition phase takes place during the last week of the semester, which typically includes the final project demonstration to the project partner and the entire class. Larger development projects are sometimes distributed over multiple terms, with handoffs to different student teams working on different aspects of the project \cite{Barker, Kurkovsky2024}. 

\subsection{Impact on Community Partners}

Service learning researchers argue that successful service learning projects embrace service learning as social justice reoriented towards the community recipients.\cite{Connolly}. Relationships are central to service learning experiences, to bolster the development of the service learning product, facilitate knowledge transfer among stakeholders, and foster transformative growth for each stakeholder\cite{Robledo, Choudhary,Soto}.  Jordaan and Mennega \cite{Jordaan} found that community partners are not just passive receivers of the benefits of service learning, but that community partners view their role as being integral in helping prepare students for the real world.  Community partners who worked directly with students reported valuing the opportunity to teach students about their organization and how it serves the community and found that the service learning experience helped them grow, specifically by being able to obtain a “fresh perspective” on their work\cite{Robledo}.  The attitude of the partner and the effectiveness of the communication has a huge impact on the quality of the service \cite{Carter}. 

Harrell et al. \cite{Harrell} conducted a survey of computing faculty. They found that student goals are often seen as more important than partner goals with about 25\% of respondents seeing benefits to partners as only a bonus \cite{Harrell}.  They also found that some programs have difficulty finding or keeping partners, as community partners potentially feel taken advantage of; while students tacitly learn that community organizations’ needs are secondary. The literature review of Yamamoto et al. \cite{Barker}, found only five projects reported using feedback from community partners to evaluate the students’ performance and and very few projects assessed the community partner’s experience, using only informal feedback if any.

It is well known that the real costs of any software development project is not in its initial development but in its ongoing maintenance and support \cite{Connolly}. Many community partners who participate in service learning lack resources for maintenance or updates %and students are not obligated to help with ongoing maintenance after the conclusion of the course 
\cite{Barker, Kilkenny}. Thus, service learning projects can actually drain the organization's resources \cite{Connolly}. It is critical that nonprofits have some means to maintain the system \cite{Bloomfield}. Research recommend that instructors help community partners and students think about long-term maintenance, or how the project can continue evolving after the student completes the course \cite{burns, MacKellar2013, Barker}. Various suggestions include finding a skilled volunteer (person or company) in the community who is willing to take over this task; or asking the students themselves to allow themselves to be contacted again to help maintain the system\cite{Bloomfield}. These time consuming tasks often fall on the instructor's shoulders, especially at small schools.

\subsection{Purpose}
%The paper should include a description of what is unique about the innovative practice, how the innovative practice differs from and builds on previous practice as documented in the literature, and new ideas that conference participants would take away from this paper. Additionally, any assessment results or relevant literature to evaluate or support the effectiveness of the innovation should be included. Authors should discuss any limitations on the transferability of the practice.
 
%Computing programs at smaller rural schools are especially challenged with how to hand-off service learning project to the community partners that do not have the expertise or resources to review and update a project on their own.
The purpose of this research is to demonstrate the use of iterative service learning, as an approach to the development of legitimately useful products for small rural organizations, in a manner that is feasible for implementation by individual faculty members at small schools with limited resources, while addressing many of the issues listed above. The goals of this innovative practice include:
\begin{itemize}
\item \textbf{Goal 1: Improving and enhancing student learning.} %discussion: The development of a social networking plan alone provides many benefits to students’ education. However, by working with members of the community through the incorporation of a service-learning initiative, it is believed students will be more enthusiastic and motivated in the development of an effective social networking plan. Motivated students, no matter their knowledge, skills, or opportunities, are more likely to succeed and experience a more enriched, higher level of learning. Instructors also benefit when teaching motivated, passionate students. 
\item \textbf{Goal 2: Furthering institutional and departmental goals toward institutionalization of civic engagement and service learning.} 
\item \textbf{Goal 3: Addressing community needs and enhancing partnerships.} %Providing interventions to manage the implementation of social networking plans at the end of the semester is critical for helping organizations effectively utilize their social networking plans. 

\item \textbf{Goal 4: Improve effectiveness of implementing social media plans for small, rural organizations.} %The goal of the iterative approach is to provide an opportunity to revise the original social networking plan to evaluate the impact of interventions in increasing implementation and sustained use rates of social media plans developed through service learning. %and research ways to make the ocial Networking capstone project a sustainable service learning tool for the foreseeable future. 
%This project also aids in the understanding of the implementation of social networking plans for rural businesses and local non-profit organizations everywhere. In general, we hypothesize that the interventions introduced will increase implementation rates, social media use, and long-term success for organizations in rural communities.
\item \textbf{Goal 5: Advancing the field of civic engagement and service learning as the pedagogy of engagement.} %This program provides an iterative mechanism to explore, develop, and evaluate new strategies regarding service-learning initiatives.
\end{itemize}

\section{Methodology}

\subsection{Program Details}

Social media plans are vital for organizations to fully engage with their audiences, but most most small rural organizations  do not have the expertise or resources to implement one. Yet by not doing so, they limit their organizational reach. This project added a service learning component to an asynchronously offered online Social Networking  capstone course %in the Department of Cyber Systems at the University of Nebraska at Kearney (UNK),
at a regional rural university. The Social Networking capstone course aims to examine a cross-section of social networking information technologies encompassing textual, aural, and visual methods along with further examination regarding the effect of personal and professional interactions \cite{WEITLHARMS2024USI}. 

Prior to 2016, a traditional course setup was implemented. Student teams in the Social Networking course completed social media plans for local area businesses and non-profits, but student teams found their own client, and the projects were not coordinated across the department as service learning. Unless the students had a personal connection with the organization, once the class was over, the plan was rarely put into action. 

Since 2016, the capstone project incorporated into the Social Networking course implements aspects of team-based learning and service-learning; stimulating students to work in small groups to develop a social networking plan for a non-profit organization or local business. Upon completion of the capstone project for the Social Networking course, students have successfully written a thorough social networking plan for the organization. The plans created by students are based on best social media practices learned throughout the duration of the course. The student team act as a social media consulting group, and students are expected to meet with a representative of the organization to determine the organization’s needs. Furthermore, students are required to evaluate the organization’s effectiveness and current social media presence. Therefore, the social networking plan developed by students can be targeted to meeting the specified needs of the organization, directing the desired message of the organization to the identified target audience, and attaining the organization’s specified goals, while presenting ideas as to how to change and improve the organization's presence. The plan is required to include identified audiences, measurable goals, and a minimum of three recommended social media strategies and tactics to reach the identified goals. Additionally, students are required to collaboratively present the social networking plan they have created to a representative of the organization, highlighting new ideas and tactics the organization could utilize to improve its social media presence. 
 
For students to complete the capstone project within the time frame designated by the academic semester, several deadlines are used to help students better manage and utilize their time. By the fifth week of the semester, students are required to submit a capstone project proposal, outlining the organization each student team plans to work with along with the overarching goals each student team wishes to accomplish by the end of the semester. Sometime mid-semester the teams are required to meet with the instructor. Of course, the teams are expected to meet with their organization representatives often enough to complete their tasks. During the thirteenth week of the course, student teams must submit their drafted social networking plan to the instructor of the course for review. The instructor then provides extensive feedback regarding the prototype paper submitted by team. The feedback provided should be thoroughly considered by students when making their own adjustments to the final social networking plan they intend to present to the client organization. 

During the final week of the course, students submit a presentation of their social networking plan that is visually oriented, cohesive, and dynamic, to both the instructor of the course as well as the client organization. Students are encouraged to record and share their social networking presentations using streaming services. However, students may also provide a live presentation or meet with the instructor and representative of the organization in person.Students also complete an individual reflection regarding the social networking plan their student team created for the capstone project and a team evaluation. The project reflection should encompass the student’s beliefs centered around whether the organization’s needs were met, the benefits the student’s services provided to the organization, whether or not the organization plans to implement the recommended social media tactics and strategies, how the project related to the learning goals of the course, what the student learned through the completion of the project, how the project applies to the student’s future, and what the student learned about him/herself by completing the capstone project. 

Note, students are not authorized to ever post content on their organization’s behalf. While student teams created a social networking plan for the organization, it is the organization’s responsibility to execute the ideas present within the resulting paper and presentation, once the course is complete at the end of the semester. 
%Consequently, the service-learning opportunity present within the course were rarely fully fulfilled as little benefit transpired for the community organization partners.\cite{WEITLHARMS2024USI}

%The research goals of this project are to study the impact of the interventions in increasing implementation and sustained use rates of social media plans developed through service learning. The projects completed in previous semesters will be used as a baseline to quantitatively measure the impacts. Quantitative analysis will involve surveys with organizations, to gage the added value to their organization. Data will be evaluated and training materials and implementation procedures adjusted accordingly.

During the first iteration of this project, a university-sponsored %Rural Futures Initiative 
grant was utilized (2016-2018) to provide funding to assist with marketing the project, developing a website of training materials, and conducting training sessions. The %Buffalo County 
local-area economic development council  assisted with informing local small business and non-profit organizations about the opportunity, and encourage applications. %UNK 
Students developed online training materials for commonly used  social networking platforms with a focus on workable solutions for local organizations. These training materials were revised as needed over the life of the grant. The %UNK 
instructors teaching the course assisted with the coordination and data collection, within the individual sections of the course.  Using the grant funding, training sessions for the organizations in the program were held at the end of each semester and students provided tutoring and support to the organizations during the first six months of implementation.

\subsection{Data Collection}
\subsubsection{First Iteration}
In the first semester of this project, fall 2016, 21 small businesses and local non-profit organizations were provided with social media marketing plans. One business requested a “do-over”, and participated again the next semester. In the spring 2017 semester, 18 organizations were provided with social media marking plans. All 18 implemented at least part of their plans. In summer 2017, two sections of the course participated in this project. One course section finished in June, and worked with five organizations. The second course section finished in July, and worked with three organizations. In fall 2017, two sections of the course participated, and nine organizations participated. Eight of those organizations implemented their plans, one organization requested a “do-over” and participated in spring 2018. During the first iteration’s final semester, spring 2018, five organizations participated in one section of the course, in addition to the one do-over organization. During the entire time period of the first iteration of this project, the organizations were provided assistance from student Social Media Specialists hired through the grant and with the resources on the project’s website.

Over two hundred and fifty students were provided the opportunity to work in teams to create social media plans as a service-learning project through their class capstone assignment in this iteration of project. Over the course of this phase, three instructors taught a total of 11 sections of the course. Each organizational project was assigned to a team of 1-5 students, by the instructors. %Students were expected to reach out to the organizations for information, assess the organizations’ current situation, needs, and resources. Organizations were expected to provide the requested information. At the end of the semester, the organizations were provided with a written social media plan, as well as a presentation or video of the students explaining the social media plan.
As part of the grant funding, organizations were also provided with a beginning and end of semester meeting, explaining the project. During the life of the project, the organizations also had access to online resources and student social media tutors to train/advise them for the first six months of implementation. However, organizations were expected to do the implementation themselves. 

\subsubsection{Second Iteration}

This research also provides an analysis of organizations’ changing impressions of social media and online networking overtime. In the 2023, due to curricular changes, only one section of the Social Networking course was offered, with 37 students. These students were assigned into teams to work with nine selected local organizations (five small businesses and four local non-profit organizations) who partook in the capstone project during the initial study period (2016-2018).  The social networking plans developed years prior were reviewed and revised by current students to provide local organizations with an updated social networking plan that corresponds to the modern times of the world. The attitudes of businesses toward social media prior to and following the development and implementation of social networking plans are analyzed across the current and previous semesters. Consequently, changes in businesses’ attitudes toward social media are analyzed following the initial implementation of a social networking plan in addition to the following years of use. Upon implementation of the social networking project, quantitative analysis was conducted of the impact of the interventions of increasing implementation and sustained use rates of the social media plans developed. 

%Nine clients were selected to participate in a follow-up iterative process, where new student teams conducted client interviews, reviewed the initial plans, and analyzed metrics from the social media strategies and tactics already in place to provide updated, improved artifacts. Due to the phasing out of this course in the curriculum, the second iteration included a single section of the course, with  

%. Organizations that were assisted by student teams include the  five local small businses and four local non-profit organiations. Kearney Catholic High School Foundation,Indulge Salon, the Kearney Area Children’s Museum, Buffalo County Historical Society, and the Archway. 

%All findings should be supported by data and analysis or carefully documented observations. For this type of paper, reviewers will expect to see an analysis of one or more educational practices, including: teaching approaches; uses of instructional technologies; and/or institutional strategies to support student success. It should include design rationale, assessment methods, evidence of effectiveness, and/or achievement of desired outcomes.

\section{Results}
\subsection{First Iteration}
This paper builds on previous research which reported pre-assessment attitudes towards social media for small rural organizations who sought assistance in developing social media plans \cite{WEITLHARMS2024USI}. In the first iteration, 49 of the 60 participating organizations completed the pre-assessment survey.  The organizations surveyed saw the potential in social media, as 94\% found  social media added value to engaging customers; 86\% reported it is a necessity for reaching customers; and 84\% recognized time spend managing social media as time well spent. However, businesses also sometimes lack the time or knowledge to use social media effectively in a business setting, as only 51\% used social media channels to conduct business and 47\% used social media for gaining customer feedback. The previous research also reported initial social media usage as well as the organizations' perception on using social media for marketing. The survey asked the organization representative to select five terms from positive, neutral, and negative terms \cite{Veral}, based on the Microsoft Product Desirability Toolkit (PDT)\cite{Benedek}. The two most common words chosen were \textbf{time-consuming} and \textbf{necessary} \cite{WEITLHARMS2024USI}.

Post assessment results show that of the 60 organizations in the first iteration of the project, two requested and were allowed to re-do their social media plan. Post assessment results show that overwhelmingly, the organizations expressed satisfaction with their plans, while often providing suggestions to tweak the project going forward. 

During the initial phase of the project, the organizations were aided from student Social Media Specialists hired through the grant funding and with the resources on the project’s website. Example comments from the project’s Social Media Specialists regarding interactions with the businesses include: \emph{participant said she was impressed with the suggestions and appreciated the insights. She was glad to have this plan for a starting point.}; \emph{Participant said he was really impressed with his plan. He said there was a lot of good insight and feedback that he looks to use.}; and \emph{Participant was very happy with her plan, and said the students met with her multiple times. She mentioned she has gained followers in result of implementation.}

Organizations were provided the opportunity to evaluate the student project, and instructors and students were afforded the opportunity to provide reflections on the project. All of the organizations who participated in the post-survey found that the student(s) effectively analysed their social media marketing needs and determined the requirements for a social media plan. All but one of the organizations found that the students designed a suitable social media plan for the organization. Organizations found that students provided prompt attendance, had professional interactions and professional attitudes, understood their needs and background, fulfilled expectations, were reliable and responsible, were open and responsive to suggestions, constructive criticism, and were a pleasure to work with. A few organizations did not understand the project parameters and stated that they expected the students to help them implement the plan after presenting it.

Over time, organizations expressed desire for assistance in revising their original artifact, as expected \cite{Connolly,Kilkenny,MacKellar}. For example, one organization noted, \emph{It would be fantastic to get some new insight on our social media platforms as we are continually growing. We use social media for 95\% of our marketing because 1) it's free and we have a small budget as a non-profit and 2) we seem to reach a lot of people that way. We love working with the university on any project that we can.}	Another noted, \emph{			
%business: (Long):
Social media is prevalent with a lot of our customers, so we need to keep up. Everything we know is self-taught, so we know we are missing out by not having a social media presence, but we don't know where to start. We have a website that never changes.}

\subsection{Second Iteration}
\subsubsection{Pre-Assessment Survey}
Seven of the nine organizations participated in the pre-assessment survey of their current social media practices. All seven used Facebook, with four having more than 200 followers; four used Instragram, with three of these having over 200 followers; two used Twitter/X, and each had more than 200 followers; and two used YouTube. All but one organization reported posting at least 1-5 times per week on a social media platform, with three reporting posting 6-10 times per week. Two organizations reported posting on two different platforms each week, one reported posting on three platforms each week, and one reported posting on five different platforms each week. 

As done in the first iteration, the organization’s perception on the use of social media was evaluated by selecting five terms based on the Microsoft Product Desirability Toolkit (PDT)\cite{Benedek}. The average of all words selected, using the means from \cite{Veral} was neutral (3.08 on a 1 to 5 scale). Claude GPT \cite{Claude} summarized the average sentiment as neutral (.55 on a scale 0 to 1 scale) as well. 
Four of the seven organizations selected \textbf{Tedious}, while several words were selected by two of the seven organizations, including \textbf{Appealing}, \textbf{Influential}, \textbf{Frustrating}, \textbf{Necessary}, \textbf{Challenging}, \textbf{Overwhelming}, and \textbf{Time-consuming}. When listing reasons that tedious was selected, organizations stated, \emph{The method I use to get photos on Facebook is tedious.}; \emph{Takes awhile to figure out how to make the posts and sometimes I get frustrated and just stop.}; \emph{When I try to do things on Facebook it's hard to do.}; and \emph{takes more time then I feel I have to give.} 

A sampling of feedback on other negative terms selected includes \textbf{overwhelming}: \emph{there are too many social media sites available and which ones would best reach my customer} and \emph{Everything changes too often}; %\textbf{frustrating}: \emph{We have placed ads on FaceBook, and I never know what the reach is.} and \emph{Not overly computer savvy so concerned about my approach}; 
\textbf{intimidating}: \emph{Being new to posts I'm always nervous about pushing the "post" button...and I'm not sure I know how to use the platforms most effectively};% \textbf{challenging}: \emph{"There are multiple people posting and no clear plan for how to use what platform. I am not super experienced with posting or the best ways to use these platforms most effectively.} and \emph{Not sure how to market my business}; 
and \textbf{complex}: \emph{It's hard to figure out how to do things and then they change it when I do.}

A sampling of feedback on positive terms selected includes \textbf{appealing}: \emph{"Our services are very good} and \emph{I do enjoy the idea of managing social}; %and \textbf{Influential}: \emph{We do for social cause as well} and \emph{Lots of customers use FaceBook and I would like to target customer base}; 
and \textbf{valuable}: \emph{The ROI is good} and \emph{it is valuable to our organization to use social  to spread our mission and share our programs and activities.}

A sampling of feedback on neutral terms selected includes \textbf{necessary}: \emph{most people use social media, and don't read newspapers} and \emph{I feel it is necessary for a business to engage and communicate through social media}.

 All organizations responded in agreement that \emph{Social media adds value to engaging our customers}; \emph{Using social media is a necessity for reaching our customers}. All but one organization found \emph{Social media planning} and \emph{implementing and managing} is time well spent. Four organizations \emph{use our social media channels to get feedback from our customers}, \emph{use social media to conduct business}, and 
\emph{use our social media channels to raise awareness about our products/causes}.

When asked how the organization used social media, sample organization responses include: \emph{we buy some ads during the season of the year when customers are looking for our business}; \emph{  We used to spend lots of \$ on print advertising, and FaceBook reaches more people for less money}; and \emph{We focus heavily on Facebook right now but need to get more followers on other channels.}

When asked how the organization would like to improve, sample responses include: \emph{we need more engaging content}; \emph{help to develop a strong presence in this area}; \emph{A plan would be a great place to start...clarity about who uses what platforms and for what purpose}, and \emph{intentionality}.

\subsubsection{Mid-Assessment}
As part of the iterative process, the student teams were provided with the organization's original social media plan, and the organizational contact information. Students were also asked to review the organizations' current social media presence and provide feedback.

Sample feedback included: 
%archway
\emph{[The organization] claims to primarily promote their events to older adults, which is why they use Facebook as their primary social media. The audience that [the organization] is working to target includes the people of [local city]. The audience they are trying to reach on their Instagram page is families who have young kids. %or kids who have an Instagram account. 
%The client wanted a plan to help their business grow its social media presence so they could convert more users into customers. %and increase the reach of the museum so they can, as a result, increase the number of visitors. The museum’s problem is that during the off-season, the museum does not get that many visitors. %Only during the summer season does the museum have enough visitors to cover the bills and keep the doors open. 
They hope that with a better social media presence, they can reach more people and have a more consistent number of visitors.} 

%KACM
\emph{I think that initially when we met with our client, her problem was that she would be active on social media but getting engagement was hard for her. Facebook brings awareness to different events that could be happening within the organization and I do feel that our client has posted consistently over time. What she wants is to get more engagement on the post themselves. In order for this to happen the main thing we want to focus on is creating a visually more engaging and consistent platform by utilizing each platform to the best of its ability. Consistent posting is what this non-profit is looking to achieve, and they have achieved the ability to post consistently. %Based on the content that I see posted on their Facebook I believe it is mainly for the parents of the kids that get to visit the Children’s Museum. Facebook seems to be where they have the most followers, which means that this is where every past visitor since Facebook was created in 2006, has gone to since the start of the social media page. I see this because if I put myself in the shoes of the parents of kids that visit this museum, I would be interested in seeing pictures of my child enjoying their time at the museum. The events that are hosted are geared toward having parents and their kids do fun things together and bringing them together through interactive play. 
%For example, May 12, 2023, the [organization] % Kearney Area Children’s Museum 
%posted their announcement for the Father’s Day weekend fun golf and foot-golf tournament. The posts on their page are either to announce future events and to recognize the ones that have already happened. I think that from the Facebook page, I recognize that the [organization] %Kearney Area children’s museum 
%has been consistently posting for a while. Posting events and keeping people updated on events that have happened already keeps the community engaged.
} 

%taxidermy
%\emph{To be honest I don’t think the channel used by [organization] is effective at all. They need to post more and create more of their own content.}
%It is always hunting season for different animals and they need to be more active. They haven’t posted since January 14th, right now it is turkey season and they should be advertising their service. 

\subsubsection{Post-Assessment}

ABET accreditation Computing Accreditation Commission or CAC \cite{ABET} consideration is an important consideration in assessing learning for computing programs. ABET focuses on the extent to which student outcomes and program objectives are being met, along with proposed actions for improvement \cite{Homkes}.  The five core ABET CS student outcomes \cite{ABET} are: \emph{1. Analyze a complex computing problem and to apply principles of computing and other relevant disciplines to identify solutions; 2. Design, implement, and evaluate a computing-based solution to meet a given set of computing requirements in the context of the program’s discipline; 3. Communicate effectively in a variety of professional contexts; 4. Recognize professional responsibilities and make informed judgments in computing practice based on legal and ethical principles; and 5. Function effectively as a member or leader of a team engaged in activities appropriate to the program’s discipline.}

The organizations were asked to complete a feedback survey. Seven of the nine organizations completed the survey. Similar to \cite{Weitl-Harms}, client evaluation questions 1-6 are intuitive in how they match the ABET CS student objectives listed above, and the client evaluation question 7, where the client evaluates if they would hire the students as members of their team, partially captures ABET CS student objective 5.

%0.857142857	Analyze
%0.857142857	Design
%0.714285714	Use Tools
%0.857142857	Communicate
%0.714285714	Document
%0.857142857	Ethics
%0.857142857	Hire
%5A;.1B;1C	Grade

The client evaluation question with matching ABET CS student learning outcome\cite{ABET} and results follow. \textbf{The student(s) I worked with on the social media project:}
\begin{enumerate}
\item Effectively analyzed the problem and determined the requirements for a solution (ABET CS 1) (85.7\%)
\item	Designed a suitable solution to the problem (ABET CS 2) (85.7\%)
\item Used appropriate software design and development tools to design and develop a solution to the problem (ABET CS 2)  (71.4\%)
\item Communicated and interacted on a professional level (ABET CS 3)  (85.7\%)
\item Prepared effective documentation for both non-technical and technical software users (ABET CS 3)  (71.4\%)
\item Interacted ethically with all persons involved with the project (ABET CS 4) (85.7\%)
\item Given a suitable job opening, I would be willing to employ the student(s) with which I worked. (ABET 5) (85.7\%)
\end{enumerate}

Organizations were also asked what grade they would assign the projects. Of the seven organizations reporting, five would assign a grade of A; one would assign a B; and the other would assign a grade of C. Organizations were also asked to comment on the experience. These comments were overwhelmingly positive. For example, \emph{The students were professional and did a great job with the assignment.}, and provided \emph{informative feedback and action plan}. 

Students were asked to complete a self-reflection of the project, which are important to student empowerment, a key pillar for experiential learning \cite{Kissel,Weitl-Harms}. Student reflections were overwhelmingly positive regarding the experience. 

%cut out half of these!!!!
%%%%%%%%%%%%%%
Sample student reflections include:
\emph{I learned a lot about how to make a company's social media effective, what social media is more beneficial for what businesses, and how to tailor a channel to a specific business's needs. As someone who doesn't use social media at all, I benefited a lot from seeing just how important social media is in today's world to sustaining and building a business. If I were to ever want to create my own business in the future, I now know the benefit of utilizing social media for more than just marketing, but also maintaining engagement and retaining customers.}

\emph{I learned through this assignment that each person has a different approach to how they would resolve the exact same question, specifically how to manage different social medias as opposed to treating them all the same.}

\emph{Throughout the completion of this project, I learned the importance of creating a strategy for a social media plan. All the research I did about creating a social media plan pointed to the need to develop measurable and attainable goals. Every social media strategy requires metrics that can be analyzed to measure the success of any social plan. I learned that this is important because not every plan will be successful from the beginning, so measurement and adjustment are necessary to develop a successful strategy.}

\emph{Through our services, we developed a cohesive theme across the three social media platforms aimed to target the specified target audience of students, alumni, and potential donors. Furthermore, the content we developed focused on educating, promoting events, and fostering connections between the [organization and the local community] %KCHSF and the Kearney community. 
I learned that I do possess an interest in developing my online communication skills.}

%\emph{This project relates to the learning goals of this course because a client's needs, target audience, and goals must be determined in order to create a system plan that works to meet those goals while optimizing a client's capabilities and existing audience members. I learned that I like to figure out what the goals for a client are so that I can create content and a plan to satisfy those goals. I like to figure out why things are the way they are so I can create the most effective solution. This will apply pretty directly to my future.}

%\emph{Overall, this project was a great insight into what it would be like to improve a client's help if they needed improvement on their social media.}

\emph{Whether they actually implement the ideas or not, it gives a local business owner a different perspective of social media and how to expand their business with it. It also helped with the course because it gives us a real case to analyze and implement what we have learned. We also have to do outside research to learn more about social media marketing.}

\emph{The plan that we created offered guidance and advice to help create effective social media channels including Instagram, Facebook, and YouTube. We were able to use knowledge that we learned in class to suggest how to use the best practices of each channel and create a successful page to the best of our abilities. We were also able to relate course knowledge by suggesting the best ways to monitor the progress of the pages. The project was strongly related to the goal of learning how social networking influences people and businesses. While doing the project I learned to have good communication skills along with working in a group of people that I was unfamiliar with.}

\emph{The social media plan that I helped to put together along with my team should help to elevate all their mentioned social media platforms (Facebook, Instagram, and Twitter) if they are posted consistently. I believe that we did a good job expressing what we thought could help the [organization] %Archway
reach more people and to also bring in more people to the actual site. We all benefited from going through the process of learning how to put together a social media plan like this and the [organization] %Archway
benefited by getting a different point of view that could help them. This project helped me to learn how local organizations interact with social media and how that can be improved. %I also learned that I do not want currently to be a dedicated social media manager. %Although, 
In the future I could use some of the information gained to work towards getting more people to interact with future organizations that I am part of.}

\emph{I believe the clients needs were met. As a team we came up with goals that could could help [the organization] %Induge
develop more clientele, become more active, influence and build trust. Each member came up with separate goals and objectives for the channels which gave each channel its own unique representation. Each platform was given a plan tailored to the needs of the salon to be effective and efficient. %Understanding how Twitter worked I believe I could give useful insight on how to achieve the goals we set out through the platform. 
After consultation and observing strengths and weaknesses, I created a goals that could improve social media presence and add value to the [organization].%salon.
}
%The project showed me that a social media plan needs to be thorough and each department has to be on board with a central idea. When a social media plan is consistent so too will be the fruits of its labor. I learned that I think differently than others and sometimes my thoughts lead to different ideas. The best thing I learned is to be comfortable and prepare to be uncomfortable. This will help me in the future by giving me understanding of how important communication is when working on group projects. Great communication can provide the cohesion needed to achieve the goals we set.}

\emph{I believe my team members and I did an effective job developing a functional, sustainable social networking plan for the organization %Kearney Catholic High School Foundation (KCHSF). 
When meeting with the client, we acquired information about the organization as well as the issues and needs the client wished to address and resolve with the development of a social networking plan. As a result, my group and I had a focus to ground and guide us while designing the three different social media platforms for the organization. %KCHSF. 
}
%Through our services, we developed a cohesive theme across the three social media platforms aimed to target the specified target audience of students, alumni, and potential donors. Furthermore, the content we developed focused on educating, promoting events, and fostering connections between the [organization and local] %KCHSF and the Kearney
%community. As a student, the Capstone project was a valuable assignment that provided me the opportunity to apply the many concepts I’ve learned throughout the %CYBR-388 
%Social Networking course to a real-world situation. Furthermore, the Capstone project was highly reflective of the learning objectives of the course as to successfully develop a social networking plan I had to utilize knowledge related to the role of social media in today’s society, online communication skills, and influential marketing methods. Through this course, I learned that I do possess an interest in developing my online communication skills. Additionally, I've learned that online communication is beneficial beyond the scope of online social media marketing degrees. In my future career in the health science field, I will utilize the skills I've learned in this course to effectively communicate information online, ensuring members of the target audience receive the information presented.}

\emph{I enjoyed the real-life application of this project. I feel like many classes have final projects that are simply busy work for the students, but this project had actual needs, deadlines, and impacts on others, which was a little scary and stressful, but essential to learn for the real world after college. I thought our project outcome was exactly what the client was looking for and I believe that we set up an excellent social media schedule for them to follow with plenty of ideas on what to post and how to attract a larger audience. Overall, this was one of the best school projects that I've ever done.}%, even if I don't plan on ever becoming a social media manager.}

%\emph{With our plan we were able to ensure success and contribute to the overall growth and success of [the organization's] %Suite Child’s
%business. Looking ahead, the completion of this project allowed for me to take a closer look into the world of social media. I can use the information I learned and apply it to my future as I am now aware of how to utilize social media as a tool that can lead towards success in various different aspects.}

\section{Discussion}

The following initiatives were used to implement iterative service learning:  work with local businesses and organization who previously participated in the capstone project who are interested in receiving revised social media plan artifacts; have student teams analyse the previously completed capstone projects, assess current social media practices of the organization, and meet with the organization to develop a second iteration of the computing artifact; and implement and review surveys completed by local businesses and organizations as well as the students involved to gauge the added value of social media plans to their business or organization and to the student's educational experiences. 

While goal 1 \emph{improving and enhancing student learning} is hard to quantify, a qualitative review of the student reflections (some included above) and the team evaluations for the second iteration showed enthusiastic and motivated students. %even when the student did not see the course as directly related to their career goals. 
The development of a social networking plan alone provides many benefits to students’ education. Additionally, by working with members of the community through the incorporation of a service-learning initiative, students learning was enhanced by working with real clients to implement  effective social networking plan. The student reflections indicate that students gained confidence and felt pride in helping meet a community need. Motivated students, no matter their knowledge, skills, or opportunities, are more likely to succeed and experience a more enriched, higher level of learning. Instructors also benefit when teaching motivated, passionate students. 

Goal 2, \emph{furthering institutional and departmental goals toward institutionalization of civic engagement and service learning} was also addressed.  This project augmented the department’s service learning and community engagement activities by providing access to quality information technology education, building, and sustaining educational and economic opportunities, and promoting a higher quality of life for students. Instructors also benefit when the outcomes of their teachings lead to lasting, palpable impacts on both their students and their community.  

Goal 3, \emph{addressing community needs and enhancing partnerships} was met by serving multiple organizations each year. Each capstone project generated a lasting impact on a rural business or non-profit organizations, aiding in their overall success and reach. 

% By serving numerous organizations per year, this project should have a tremendous lasting impact on small business and non-profit success in Central Nebraska.

For the first iteration, providing grant-funded interventions to manage the implementation of social networking plans at the end of the semester was critical for helping organizations effectively utilize their social networking plans. However, this was not sustainable without funding. 

The addition of the iterative approach addressed the challenge of maintaining of the computing artifact by community partners who do not have the resources to do it in house and without outside grant funding for the university.

Goal 4  \emph{improve effectiveness of implementing social media plans for small, rural organizations} was addressed with the iterative approach. The goal of the iterative approach is to provide an opportunity to revise the original social networking plan to evaluate the impact of interventions in increasing implementation and sustained use rates of social media plans developed through service learning. %and research ways to make the ocial Networking capstone project a sustainable service learning tool for the foreseeable future. 

This project also aids in the understanding of the implementation of social networking plans for rural businesses and local non-profit organizations everywhere. In general, we hypothesize that the interventions introduced will increase implementation rates, social media use, and long-term success for organizations in rural communities.

This program also addressed goal 5\emph{advancing the field of civic engagement and service learning as the pedagogy of engagement} by providing an iterative mechanism to explore, develop, and evaluate new strategies regarding service-learning initiatives. It also utilized client evaluations and student reflections to evaluate student learning.

While this paper describes only two iterations of service learning, with only nine selected organizations in the second iteration from the original 60 organizations originally studied, further iterations are feasible and would follow the same pattern as the second iteration.

\subsection{Overall Issues}
As with any project, there were communication issues. One student commented, \emph{When we are assigned a client who does not respond to emails after 3 weeks, isn't there to answer questions on multiple occasions, and then when she is there, she directs me to talk to someone else who does not even work for her business, it makes it hard to understand what they want from our group.} One organization commented, \emph{The students gave some good ideas but we are already implementing what they said. Still good ideas though.} Another organization commented, \emph{I believe this project was thrown together last minute and not as comprehensive as this assignment should have been being a capstone course.} The instructor served as a facilitator to clear communication issues as they arose. 

Each section of this course was offered asynchronously online. This did not seem to have a negative impact on the results, as one student noted, \emph{This was a very interesting project to do, especially for an online class and having to coordinate with the other members since we did not have the opportunity to do so in class.} Instructors in the first iteration addressed some early communication issues due to the asynchronous online nature of the class. One adjustment was the inclusion of Zoom to help the students get to know the instructor and course expectations better, as well adding a required mid-semester team meeting with the instructor. 

\subsection{Transferability}
The innovative practice described in this paper is transferable directly for service learning projects utilizing social media plans as the computing artifact. Social media plans are vital for organizations to fully engage with their audiences, and the plans and social media activity should be regularly reviewed. Since most most small organizations do not have the expertise or resources to implement one, iterative service learning is a highly effective mechanism.

For computing artifacts such as software or websites, this innovative practice would require that the student teams have access to the working code, which can sometimes be an issue with changing technologies. Additionally, code bases often take a much longer time to review and understand as compared with social media plans. However, the concept of using iterative service learning, with teams of students, even five or more years later, to provide necessary maintenance on a service learning computing artifact is a viable alternative. Other transferability ideas demonstrated in this project include incorporating client evaluations (based on ABET learning outcomes) and student reflections into the grading.

Finally, in many situations, iterative service learning is likely a more viable approach than finding a skilled volunteer (person or company) in the community willing to take over this task; or asking the students themselves to allow themselves to be contacted again to help maintain the system as suggested in \cite{Bloomfield}, especially if the updates requested are not immediately after the initial hand-off, but years later.

%improvements?
%The paper should include a description of what is unique about the innovative practice, how the innovative practice differs from and builds on previous practice as documented in the literature, and new ideas that conference participants would take away from this paper. Additionally, any assessment results or relevant literature to evaluate or support the effectiveness of the innovation should be included. Authors should discuss any limitations on the transferability of the practice.

\section{Conclusions and Future Work}
Future work includes conducting and evaluating various service learning approaches, including further iterations of the iterative approach introduced here. Future work also includes a longitudinal impact study of the interventions in increasing implementation and sustained use rates of social media plans for the organizations served over the past ten years through service learning in this course, to compare organizations who were served in the various iterations of the course; as well as to compare with organizations in the local area who did not work with the university students. Future work includes conducting iterative service learning in other computing courses, particularly in courses where the artifact is software or websites.  

This research introduces an iterative approach to service learning to develop, review, and  improve computing-based artifacts for small rural organizations, over an extended period. This approach addresses the well-known issue that plagues the success of service learning in computing programs; the fact that computing artifacts rarely function well long-term without versioning and updates. The fact that service-learning projects are often one-time engagements, completed by single teams of students over the course of a semester or year long course limits the benefit for the community partners, such as small rural organizations, that do not have the expertise or resources to review and update a project on their own.

The innovative practice of using iterative service learning was demonstrated through the development of social media plans for small rural organizations through service learning over several years. Teams of undergraduate students in a computing capstone social media development course created tailored social media plans for numerous small rural organizations. The approach was demonstrated with an second iteration using nine selected clients who had previously received a social media plan as part of the service learning five years earlier. The new student teams conducted client interviews, reviewed the initial plans, and analyzed metrics from the social media strategies and tactics already in place to provide updated, improved artifacts. Using ABET computing learning objectives as a basis, clients reviewed the student teams and the artifacts created. Students also reflected on their experiences. Overwhelmingly, both the client and student feedback from this approach were positive.  This research demonstrates an innovative practice for creating and maintaining computing artifacts through iterative service learning, while addressing the resource constraints of small rural organizations.

\bibliographystyle{IEEEtran}
\bibliography{BibFile}

\end{document}